\documentstyle[preprint,aps]{revtex}
\begin{document}
\title{Stability of Elastic Glass Phases in Random
Field XY Magnets and Vortex Lattices in Type II
Superconductors}

\author{Daniel S. Fisher}

\address{Department of Physics, Harvard University,
Cambridge, MA 02138}

\maketitle
\begin{abstract}
A description of a dislocation-free elastic glass phase
in terms of domain walls is developed and used as the
basis of a renormalization group analysis of the
energetics of dislocation loops added to the system. 
It is found that even after optimizing over possible
paths of large dislocation loops, their energy is still
very likely to be positive when the dislocation core
energy is large.  This implies the existence of an
equilibrium  elastic glass phase in three dimensional
random field X-Y magnets, and a dislocation free,
bond-orientationally ordered ``Bragg glass'' phase of
vortices in dirty Type II superconductors.
\end{abstract}


It has been believed for a long time that systems with
quenched randomness that couple to a continuous symmetry
order parameter cannot exhibit long range order in less
than four dimensions.
\cite{lar1}-\cite{mar4}  Originally it
was argued that positional order of Abrikosov vortex
lattices in superconductors would be destroyed by random
pinning. \cite{lar1} Recently for a class of random
systems including X-Y magnets in a random magnetic field,
\cite{imr2} the absence of long range order has been
proven rigorously. \cite{aiz3} Yet an intriguing open
question remains: for weak randomness in such systems,
are they simply disordered at low temperatures or can
phases exist which exhibit some kind of topological or
other type of order that distinguishes them from high
temperature disordered phases? This issue has resurfaced
in the context of vortices in high temperature
superconductors; various authors have either implicitly
assumed, raised the question of, or given qualitative
arguments for, the existence of an {\it elastic vortex
glass} phase which is locally lattice like and is free
of large dislocation loops. \cite{bla5}-\cite{gia7} Such
a phase would probably have power law Bragg-like
singularities in its structure factor and in addition,
have true bond orientational long range order,
\cite{gia7} thus providing a counter-example to the
general conjecture mentioned above. \cite{mar4}

In the simpler context of three dimensional random
field X-Y magnets, Gingras and Huse \cite{gin8} have
explicitly conjectured, and given some numerical
evidence in support of, the existence of a phase
transition to a defect-free phase for  weak-randomness 
at low temperatures.  Yet at this point, no convincing
analytical arguments to support or deny the existence of
an elastic glass phase have been put forth,
\cite{gia7},\cite{kie9} although the delicate balance
between elastic randomness and dislocation energies has
been pointed out by Giamarchi and Le Doussal \cite{gia7}.

In this paper, we explicitly study the stability of
a putative elastic glass phase in a three
dimensional random field X-Y model to dislocation
loops---to avoid confusion, we will refer to the
relevant topological defects in all these systems as
''dislocations''. To do this we must first reconsider
the behavior of the ground state and excitations of an
elastic glass model with dislocations excluded by fiat.
The framework that will be developed, naturally allows
one to analyze the energetics of a dislocation loop
that is added to the system.  If the core energy of
the dislocation line is sufficiently large, it is
found that, even after optimizing over the possible
paths of a dislocation loop, large loops cost energy
with high probability.  This situation can be achieved
for a weak random field X-Y model and we thus conclude
that an elastic glass phase {\it should exist} in this
system in three dimensions.

By analogy, our results are applied to other systems,
especially the elastic vortex glass.

Our basic starting point will be the elastic glass
model with hamiltonian

\begin{equation}
{\cal H}=\frac{1}{2}\sum_{(xy)}\left[\varphi{(x)}-
\varphi{(y)}\right]^2-h\sum_x\cos \left[\varphi{(x)}-
\gamma{(x)}\right]
\label{eq:a}
\end{equation}
with $\gamma(x)$ independent quenched random variables on
each site uniformly distributed on $[-\pi, \pi]$ and
$\varphi(x)\epsilon(-\infty, \infty)$.  This model has
been extensively studied by a variety of techniques,
including an approximate real-space renormalization
group (RG) \cite{vil10}, a perturbative $4-\epsilon$ RG
expansion \cite{gia7} and an approximate variational replica
symmetry breaking (RSB) calculation. \cite{gia7}  At all
temperatures the behavior is controlled by a zero
temperature fixed point whose properties yield disorder
averaged (denoted by an overbar) mean square phase
variations
\begin{equation}
{\overline{<\left[\varphi(x)-\varphi(y)\right]^2>}}
\approx 2A\ln|x-y|
\label{eq:b}
\end{equation}
at large distances, with $A$ a universal coefficient
computable in the $4-\epsilon$ expansion. \cite{gia7} It is
believed that the mean correlation function will decay as
${\overline{<e^{i\varphi(x)}e^{-i\varphi(y)}>}} \sim
\frac{1}{(x-y)^\eta}$
(although the gaussian variational RSB \cite{gia7} result that
$\eta=A$ should {\it not} be correct).  These results
have essentially been obtained from coarse graining or
Fourier space representation of the phase
variables. Unfortunately, this framework does
not appear to be naturally amenable to consideration of
dislocations, for these intrinsically involve, as we
shall see, physics on many length scales.  

A complementary and more complete picture of the
elastic glass phase can be constructed in terms of
domain walls which turn out to be the natural objects at
long length scales.  This is most easily seen by
studying the limit
$h\to\infty$ so that
$\varphi(x)=\gamma(x)+2\pi n(x)$ with $\{n(x)\}$
integers. The ground state can then be represented
[up to a uniform shift in the
$\{n(x)\}$] by the oriented surfaces through which $n(x)$
changes by $\pm 1$.  The typical $\varphi(x)
-\varphi(y)\sim\pm\sqrt{A\ln (x-y)}$ arises from a sum
of random $\pm2\pi$ terms from  crossing the nested
set of closed surfaces enclosing $x$ and $y$; these
occur typically on scales $1, B, B^2\cdots$ with
$B\sim e^{1/A}$.  In contrast, the coarse
grained
$\varphi$ averaged over regions of
size of order
$\frac{1}{2}|x-y|$ will vary from $x$ to $y$ by only
$O(1)$. 

We will primarily be interested in
studying the ground states of the system with different
boundary conditions---later including
boundary conditions induced by dislocations.  Thus rather
than working with the surfaces across which $n(x)$
changes, it is useful, as for Ising spin glasses, \cite{fis10A}
to consider configurations {\it relative} to the ground
state
$\{n_G(x)\}$ with some chosen fixed boundary conditions;
any state can then be represented by the set of {\it
oriented domain walls} across which
$n(x)-n_G(x)$ changes. A crucial question is the
typical energy of the minimal domain wall excitation,
$\varepsilon_L$, that surrounds a chosen volume $L^3$.

The observation that the coarse grained $\varphi$ has
variation of O(1) strongly suggests from the scaling
of the elastic energy in Eq.~(\ref{eq:a}), that
\begin{equation}
\varepsilon_L\sim L^\theta\ {\rm with}\ \theta=1
\end{equation}
(generally $\theta=d-2$).  This can be seen more
explicitly by fixing $\varphi(x_1=0, x_2, x_3)=0$ in a
system of size $L\times L\times L$ and letting
$\varphi(x_1=L, x_2, x_3)$ change from zero, which
defines the reference state  $\{n_G(x)\}$, to
$2\pi$, with periodic boundary conditions in $x_2\to
x_2+L$ and $x_3\to x_3+L$. This forces a single domain
wall spanning the system and changes the energy by
$E_L$.  We can now make use of the 
powerful statistical symmetry  of the model hamiltonian
Eq.~(\ref{eq:a}): if $\varphi$ is replaced by
$\varphi=\varphi_D + \psi$ with $\varphi_D$ single
valued module $2\pi$ and
$\nabla^2\varphi_D=0$ [with lattice derivative
operators] then the {\it statistical} properties of
\begin{equation}
\tilde{\cal H}(\psi)={\cal H}-
\frac{1}{2}\sum|\nabla\varphi_D|^2
\label{eq:c}
\end{equation}
with constant boundary conditions on $\psi$ are identical
to those of
${\cal H}(\varphi)$ with constant boundary conditions on
$\varphi$. Choosing $\varphi_D=\frac{2\pi x_1}{L}$, this
implies that the mean energy of the forced domain wall
is $\overline{E_L}=\frac{1}{2}(2\pi)^2L$.  The necessary
balance of the random part of the energy with the mean
elastic part $(\sim L)$, implies that this is consistent
only if the variations of the spanning wall energy,
$E_L$ are also of order $L$; i.e. $\theta=1$.  

If in the $L^3$ system, the boundary condition at
$x_1=L$ is changed to $\varphi=2\pi s$, then $s$
spanning domain walls will be forced. Because these
are closer together for larger $s$, they can less easily
optimize their positions and their energy will be
larger.  Following arguments for confined directed
polymers, \cite{fis11} we consider each section of domain wall of
scale $L/s$. Roughly, these are transversely confined by
the others on scale $L/s$.  Thus we guess that the mean
energy of a section will be of order $L/s$ yielding, with
$s^3$ sections, a total increase in the mean energy
$\overline{E_L}\sim Ls^2$ in agreement with the result
from the statistical symmetry.  The sample-to-sample
variations in $E_L$ can be guessed from a central
limit sum of $s^3$ roughly independent variations of
each section yielding
$\delta E_L\sim s (L/s)\sqrt{s^3}$. For
$s\sim\frac{L}{2}$, this yields $\delta E_L\sim
L^{\frac{3}{2}}$ as should be expected since the
resulting system's ground state---and indeed its
hamiltonian $\tilde{\cal H}$---is very different
everywhere from the $s=0$ reference system.

The main lesson from the above is that sections of
optimal domain walls on scales $L$ that are {\it confined
on the same scale} $L$, typically have energy 
that is distributed with positive mean and variations
both of order
$L$.  If we try to put several distinct domain walls
of scale $L$ into a volume of order $ L^3$, the energy
of each of them will have to increase.

We now consider inserting a single dislocation loop of
radius
$R$ into a fixed position of the system by making
$\varphi$ multivalued with $\nabla\times\nabla\varphi 
=2\pi$ on plaquettes through which the dislocation loop
passes and $\nabla\times \nabla\varphi=0$ elsewhere.  If
the core energy per unit length of the dislocation is
$\epsilon_0$, the statistical tilt symmetry implies that
the extra groundstate energy, $D_R$, due to the
dislocation has mean
\begin{equation}
\overline{D_R}=2\pi R\left[\epsilon_0+\pi\ln R \right]
\label{eq:d}
\end{equation}
with the short-scale cutoff of $\ln
R$ absorbed into $\epsilon_0$. How can we understand this
in terms of domain walls? The dislocation loop forces in
a single domain wall that spans the loop.  The sections
of the wall of scale $L=1$ adjacent to the loop are
confined on this scale and thus each have mean energy
of order 1 and roughly independent variations of this
same order. The sections of scale $L=2$ are attached to
these which confines them on scale $L=2$ and gives each
a mean energy of order 2; and so on, on scales 4, 8, 16
up to
$L\sim R$.  Thus each factor of two in scale will
contribute a factor of $R$ to $\overline {D_R}$
yielding the $R\ln R$ of Eq.~(\ref{eq:d}). But the
variations $\delta D_R$ of $D_R$ will be much smaller:
 from each scale $L$ there will be a random contribution
$\pm\left(R/L\right)^{\frac{1}{2}}L$, from the sum of
order $R/L$ roughly independent variations of sections
of scale $L$.  The typical variations in
$D_R$ are thus 
\begin{equation}
\delta D_R\sim R
\label{eq:e}
\end{equation}
which are dominated by the {\it largest} scale section
of the wall.

For a large {\it fixed} dislocation loop, the energy is
thus very likely to be large and positive.  But we must
consider the optimization of the dislocation energy over
all possible paths of the dislocation loop in a region
of volume of order $R^3$. We do this by an approximate
renormalization group analysis of the effects of
sections of the wall at each length scale  on the
optimal path with minimum energy of a segment of the
dislocation loop.  

We focus on transverse deformations
of a dislocation segment by distances of order $W$ with
the effects of smaller scale deformations of the
dislocation and the concomitant changes in the sections
of wall attached to it on scales smaller than $W$,
included in an effective mean dislocation energy per
unit length
$\tilde\epsilon_W$ with local variations around this
value.  If the dislocation were straight on smaller
scales,
$\tilde\epsilon_W$ would  be simply $\epsilon_0+\pi \ln
W$ but we expect it to be reduced from this by the
smaller scale deformations.  Our task is to
iteratively understand how the deformations on a scale
$W$ change
$\tilde\epsilon$ on larger scales.

In the continuum approximation, (valid on large scales)
the typical excess energy cost of a transverse
distortion of a segment of length $\Lambda$ of the
dislocation by an amount $W<<\Lambda$ will be
$\frac{1}{2}\tilde\epsilon_WW^2/\Lambda$ from the extra
length of the dislocation.  Such a distorted dislocation
segment will have a different spanning wall attached to
it than the straight (on scale $W$) segment.  But if this
spanning wall were completely different over the whole
length
$\Lambda$, it would imply the existence of many ($\sim
\Lambda/W$) distinct minimal walls in a volume
$\Lambda^3$ which is highly improbable by the earlier
discussion of confined walls.  What should be expected,
instead, is that the minimal spanning walls attached to
the two different paths of the dislocation will
typically only differ in a strip of width $W$ near the
dislocation, with all of them attaching to the {\it same}
optimal wall further away.  

Each section of scale $W$
will be roughly independent and the mean energy of each
of these sections of the two wall configurations will be
the same with differences between them also of order
$W$.  Thus the {\it total difference} in the attached
spanning wall energies  between the energies of the
distorted and undistorted segment will be of order
$\pm\left(\frac{\Lambda}{W}
\right)^{\frac{1}{2}}W$.  Balancing this difference
against the excess effective core energy cost of the
distortion from above, yields the typical length
$\Lambda_W$ over which transverse deformations of size
$W$ will occur:
\begin{equation}
\Lambda_W\sim W\tilde\epsilon_W^{\frac
{2}{3}}
\label{eq:f}
\end{equation}
which is $>>W$ if $\tilde\epsilon_W$ is large.

At the next length scale, $2W$, the effective
$\tilde\epsilon$ will change. The inclusion of the wall
segments of scale $W$ increases the mean energy of a
length $\Lambda_W$ segment by $\sim W(\Lambda_W/W)$ but
the optimization over the scale $W$ deformation of the
dislocation decreases it by $\sim
\left(\Lambda_W/W\right)^{\frac{1}{2}}W$. Thus we find
that 
\begin{equation}
\tilde\epsilon_{2W}\approx\tilde\epsilon_W+\pi \ln2-
\alpha/\tilde\epsilon_W^{
\frac{1}{3}}.
\label{eq:g}
\end{equation}
with $\alpha$ some coefficient.

This is our key result: although the arguments leading
to it are not precise, the {\it form} should be correct
for large $\tilde\epsilon_W$.  The RG flow of
Eq.~(\ref{eq:g})
implies that the delicate balance \cite{gia7} between the $R\ln
R$ terms in $D_R$ and its $\pm R$ variations, as well as
that implied by the almost linear growth in
Eq.~(\ref{eq:f}) of dislocation distortions $W_\Lambda$
with $\Lambda$, is resolved by the dominance of the
deterministic terms in the dislocation energy over
even the optimal random ones.  

For sufficiently large $\epsilon_0$, the renormalized
energy of the lowest energy dislocation loop of radius
$R$ in a volume $\sim R^3$ can now be obtained by
renormalizing until a scale $W_R$ at which
$\Lambda_W\sim R$. On longer scales, the optimal
dislocation loop of radius $R$ will look essentially
circular but can still reduce its energy by rotating or
moving within the region of volume $\sim R^3$. From
scales
$W$ in the range $R>W>W_R$, the renormalization of
$\tilde\epsilon$ will have a similar form to
Eq.~(\ref{eq:g}) but with the last term replaced by
$\alpha'(W/R)^{\frac{1}{2}}$. The mean energy of the
optimal dislocation loop of radius $R$ in a volume
of order $R^3$ will thus be
\begin{equation}
\overline{D_R}\approx 2\pi R\left\{\pi \ln R+\epsilon_0
-O\left[\epsilon_0^{\frac{2}{3}},\left(\ln R
\right)^{\frac{2}{3}}\right]\right\}
\label{eq:h}
\end{equation}
for large $\epsilon_0$.

The largest scale $W\sim R$ should dominate the
variations in $D_R$, yielding
\begin{equation}
\delta D_R\sim R<<\overline{D_R}.
\label{eq:i}
\end{equation}
Large dislocation loops with negative energy will thus
be very improbable. \cite{12} Some small loops
will of course appear for any $\epsilon_0$. The
effect of these in the presence of an applied
$\nabla\varphi$ will decrease the effective long
wavelength elastic constant of
the vortex glass phase and concommitantly the mean
energy of large dislocation loops by allowing spanning
walls with small holes in them.  But if $\epsilon_0$ is
greater than some critical value, $\epsilon_{0c}$, these
effects will only yield finite renormalizations and
the elastic glass phase will be stable.  In contrast,
for $\epsilon_0<\epsilon_{0c}$, $\tilde\epsilon_W$ will
decrease with length scale eventually becoming negative
and leading to the proliferation of dislocations of
size greater than a correlation length $\xi$, even at
zero temperature.  

So far, we have focused on an unphysical limit of
infinite $h$ with $\epsilon_0$ tuned by hand.  But the
results will apply much more generally.  For weak
randomness ($h<<1$) in a three dimensional X-Y magnet,
the crossover length scale
$\xi_p$ above which the random fields become important
diverges as $\xi_p\sim h^{-2}$. \cite{imr2}  On scales $L$
larger than $\xi_p$, the randomness typically prevents
the system from being able to find more than one optimal
configuration in a region: another configuration that
differed over most of a region of size $L^3$ would be
expected to differ in energy by of order
$h^4L^3$ which, since this is much larger than the
basic energy scale $L$ for excitations and stiffness, is
highly unlikely.  Thus we expect configurations with
different boundary conditions (or those differing by
large dislocation loops) to differ substantially by
other than multiples of
$2\pi$ only on lower dimensional subsets: these will be
domain walls with thickness $\sim\xi_p$.

Our analysis can be carried through for small $h$ if the
system is first coarse grained to a scale $\xi_p$. This
yields an effective core energy
$\tilde\epsilon_{\xi_p}\approx\pi\ln\xi_p$ from
distances within $\xi_p$ of the dislocation since the
energy on these scales is mostly the deterministic
elastic energy.  Larger scale deformations of a
dislocation are similar to those studied above.
If $h$ is sufficiently small,
$\tilde\epsilon_{\xi_p}$ will be large and the energy
gained from optimizing the dislocation path will be
small; the elastic glass phase will thus be {\it stable} for
weak random fields.

The static properties of the elastic glass phase 
are somewhat subtle. The expected power law decay of
$e^{i\phi}$ correlations in the model system \cite{gia7}
should also occur in the elastic glass phase of the X-Y
random field system.  The phase should also exhibit a
non-zero elastic stiffness: a {\it mean} excess energy 
cost of order
$L$ if opposite boundary conditions on $\cos \varphi$
are applied at two ends. But its primary distinguishing
features  will be {\it dynamic} as are
manifested in the analogous elastic vortex glass phase to
which we now turn.

Aside from the complications of dislocations (and domain
walls) with different Burgers' vectors and of coupling of
the real and order parameter spaces, the arguments given
above for X-Y random field magnets should apply also for
three dimensional lattices in weak random potentials
{\it both} for solids in porous media and for vortex
lattices in dirty Type II superconductors.  Power law
singularities at Bragg peak positions should be
observable in the resulting elastic  glass phases, \cite{gia7}
especially if the exponent $\eta$ is small. 
Furthermore, the absence of large scale dislocation
loops means that the bond-orientational stiffness will
{\it not} be local and, even though the randomness
couples to the local lattice orientation, long range
hexatic order should exist. [4, 7] 
But once dislocations proliferate,  as the randomness
is increased, the orientational stiffness will become
short range and the randomness coupling to local bond
orientations will make the system fully disordered.
\cite{mar4}

The dynamic properties of an ideal dislocation free
elastic vortex glass phase have been 
extensively studied. \cite{bla5}  With only small scale
dislocations present, the superconducting properties of
this phase should persist provided that the phase is
``commensurate'' with one vortex per unit cell; \cite{fre13} if
not, the dynamics of the vacancy and the interstitial
lines  \cite{fre13} will dominate the dynamics with the
remaining  vortices essentially frozen.  One might
worry that dislocation excitations rather than motion of
sections of the lattice could dominate the dynamics, but
their higher energy cost ($L\ln L$ vs. $L$) and less
likelihood of having low energy, argue against this.

For
weak pinning, a first order transition directly to the
commensurate elastic vortex glass should occur; this is
the residual of the pure system melting transition
that will occur regardless of the nature of the low
temperature phase and is consistent with experiments.
\cite{zel14}  For stronger pinning, various other phases
might exist: a superconducting inelastic vortex glass
phase with proliferated dislocations; \cite{fis6} and
incommensurate elastic glass phases that could be either
superconducting or not.  Investigation of
possible phase diagrams and transitions, as well as
development of the theoretical understanding of the
simpler X-Y systems may have to await further progress
in methods for analyzing these subtle systems.

I would like to thank David Huse, Terry Hwa and Matthew
Fisher for useful conversations.  This work is supported
in part by the NSF via DMR 9106237, 9630064 and Harvard
University's MRSEC.


\begin{references}
\bibitem{12} The distribution of
$\Gamma\equiv(D_R-\overline{D}_R)/R$ should be scale
invarient for large $R$ in the elastic glass phase.  If
Prob ($\Gamma)\sim e^{a(-\Gamma)^\alpha}$ for large
negative $\Gamma$, then one would expect a density of
dislocation loops of radius $R$ in the ground state of
order $e^{-b(\ln R)^\alpha}$, illustrating the
almost-marginal nature of the phase.
\end{references}
\end{document}